\documentclass[prd,superscriptaddress,showpacs,preprintnumbers,amsmath,amssymb]{revtex4}
\usepackage{amsmath}
\usepackage{amstext}
\usepackage{amsopn}
\usepackage{amsfonts}
\usepackage{amssymb}
\usepackage{bbm}
\usepackage{accents}
\usepackage{empheq}%for overbracket
\usepackage{graphicx}
\usepackage{epsf}
\usepackage{graphics}
\usepackage[latin1]{inputenc}

\begin{document}

\title{Geometric Tachyon and Warm Inflation}
\author{Anindita Bhattacharjee}
\affiliation{Department of Physics, Assam University, Silchar, Assam 788011, India}

\author{Atri Deshamukhya\footnote{atri.deshamukhya@gmail.com}}
\affiliation{Department of Physics, Assam University, Silchar, Assam 788011, India}

\begin{abstract}
The inflationary models developed in presence of a background radiation can be a solution to the reheating problem faced by common cold (isentropic) inflationary scenario.
A D-brane system comprising of $k$ Neuvo-Schwarz(NS) $5$-branes with a transverse circle and BPS D$3$-branes with world volume parallel to the NS 5-branes, placed at a point on the transverse circle diametrically opposite to the NS $5$ branes has a point of unstable equilibrium and the D$3$ brane has a geometric tachyonic mode associated with displacement of the brane along the circle.
Cold inflationary scenario has been studied in connection with this geometric tachyon \cite{PA1} where it was found that one needs a background of minimum $10^4$ branes  to realize a viable inflationary model.
In this piece of work, we have tried to study a model of inflation driven by this geometric tachoyn in presence of radiation. We have found that compared to the isentropic scenario, to satisfy the observational bounds, the number of background branes required in this case reduces drastically and  a viable model can be obtained with even six to seven NS $5$-branes in the background. In this context, we have also analyzed the non-gaussianity associated with the model and observed that the concerned parameter lies well within the observation limit.
\end{abstract}


\maketitle

\section{\bf Introduction}

Inflation so far seems to be the most successful proposition which can solve the shortcomings of standard Big Bang Model and also can explain the temperature an-isotropies observed in CMBR. It is a process of extreme accelerated expansion of the Universe. There are two dynamical realizations of cosmological inflation, the standard scenario \cite{GA,AS,Linde}  referred as cold inflation and warm inflation \cite{BB,BB1}.

According to the standard inflationary model the nascent universe passed through a phase of exponential expansion that was driven by a negative-pressure vacuum energy density. The high vacuum energy density drove this accelerated expansion during the very early moments of the Big Bang. This expansion is responsible for the continuing Hubble expansion of the Universe, its present size, its relative homogeneity and isotropy, and its spatial flatness. Inflation also explains the origin of the large-scale structure of the cosmos. The quantum fluctuations of the inflation field produced ripples in space-time, which are thought to be the primordial seeds of the galaxy formation.These models assumes that any radiation present in the universe prior to the inflationary period rapidly dilutes out and there is no further radiation production during inflationary era.

In warm inflation, the energy stored in vacuum is converted into radiation due to dissipation effects during the period of inflation which never leaves the universe cold in contrast to the standard cold scenario where at the end of inflationary period the universe is left at supercooled state and a mechanism of reheating is required for generation of relativistic particles and reheat the universe. The basic idea of warm inflation is that during inflationary period the inflaton field is coupled to several other fields but no apriory assumption is made regarding the multi-field interaction.That makes it explicit that the thermodynamic state of the universe during inflation is a dynamical question \cite{GB}. As the inflaton relaxes towards its minimum energy configuration, it is assumed that it decays into lighter fields, generating an effective viscosity. If this viscosity is large enough, the inflaton will reach a slow-roll regime, where its dynamics becomes over-damped. This over damping is efficient to get the correct number of e-folds before the end of inflation \cite{PA2}. The inclusion of thermal effects acts similar to a mass term which breaks the scale-symmetry of the zero-temperature theory \cite{AB2}. In Ref.\cite{BR} it has been argued that the coupling constant fine-tuning problem is closely associated to this scale symmetry and breaking of this scale symmetry can avoid this problem. Hence warm-inflation can be considered as a successful mechanism of inflation sans independent reheating period.

One of the theoretical issues with the paradigm of inflation is that in most of the models, instead of being motivated from a fundamental theory, the scalar field responsible for driving inflation is chosen ad-hoc. In addition, it is difficult to device a consistent mechanism for reheating in each model specifically when the potentials are of runaway type. The mechanism of reheating is very crucial particularly the particle production and energy transfer issues. In last decade there are many attempts to construct successful models of inflation driven by scalar fields motivated from String Theory. But the cold-inflationary models, though could account for the cosmological observables \cite{WMAP} lacks a successful reheating mechanism at the end of inflationary mechanism in many cases.

In the string theory context, one of the attractive inflationary model, that has drawn considerable attention in last few years, is where the tachyon field associated with an unstable D-brane \cite{Gib,FTCP,FTCP1} plays the role of the inflaton. If the tachyon field starts to roll down the potential, then the universe dominated by a new form of matter will smoothly evolve from inflationary universe to an era which is dominated by a non-relativistic fluid \cite{Gib}. The dynamics of the tachyon field is described by a non standard effective action \cite{Gar,G1,G2}.

Surprisingly, the motion of a D3-brane in the background of a stack of coincident NS$5$-branes(Geometric Tachyon) is also found to be described by a similar action \cite{kuta,kuta1}. In this system, the particle experiences a proper acceleration orthogonal to its proper velocity due to the background dilaton field which changes the dynamics from that of a simple geodesic motion. In particular, it is shown that in the vicinity of the five-branes, it is this acceleration which is responsible for modifying the motion of the radial mode to that of an inverted simple harmonic oscillator leading to the tachyonic instability \cite{PA3,PA4}.

This geometric tachyon field has been studied as a candidate for driving inflation successfully \cite{PA1}. It was observed that though this tachyonic field is capable of driving cold inflation successfully, one needs a large number of ($10^4$) branes to provide the necessary background. This number being a little high we wish to revisit the scenario in presence of radiation to see whether in addition to waiving out the issue of reheating that can bring change in the requirements of number of NS$5$-branes in the background.

In Section.II we discuss the formalism and in Section.III the observed result.

\section{Formalism of Warm Inflation}

In the cold inflationary scenario, the inflation dynamics is governed by the Einstein equations for FRW metric and the tachyon field equation derived from the tachyonic action (see \cite{FTCP1} for details)i.e
\begin{equation}
{\dot{\rho}_{\phi}}+3H(\rho_{\phi}+p_{\phi})=0 \nonumber
\end{equation}
where $H=\dot{a}/a$ is the Hubble parameter, $\rho_{\phi}$ is the field energy density and $\rho_{\gamma}$ is the radiation energy density.
However, in the warm inflationary scenario, i.e when thermal effects are taken into account, the corresponding equations are modified due to the contribution of radiation energy density and the inflation  dynamics is described in general by the following equations \cite{Herrera} where the dissipative effect is taken into account by introducing a term proportional to tachyon velocity.
\begin{equation}
H^2~=~\frac{1}{3 M_P^2}(\rho_{\phi} + \rho_{\gamma})
\end{equation} ,
\begin{equation}
{\dot{\rho}_{\phi}}+3H(\rho_{\phi}+p_{\phi})=-\Gamma {\dot{\phi}}^2 \nonumber
\end{equation}
where $\Gamma$ is the dissipation coefficient which
is assumed to be responsible for the decay of  tachyon field
into radiation in the inflationary regime and $M_P$ is the reduced Planck mass. In general, the dissipation coefficient is expected to be a positive function of the tachyon field. It may be worth mentioning here that attempts have been made to obtain scalar field evolution equations for warm inflation from first principles which found that dissipative effects during warm inflation are temporarily non-local although limits have also been obtained for which the effects are of the local form  \cite{GB,GAJ,GAJ1}. In the present model we consider the dissipation coefficient as an arbitrary function of tachyon field.

The energy density ${\rho}_{\phi}$ and pressure $p_{\phi}$ for
the tachyon field $\phi$ are :
\begin{eqnarray}
\rho_{\phi}~&=&~\frac{V(\phi)}{\sqrt{1-{\dot{\phi}}^2}}\\
p_{\phi}~&=&~-V(\phi)\sqrt{1- \dot{\phi}^2}
\end{eqnarray}
where $V(\phi)$ is the potential function associated with the tachyon field.
Equation (1) then takes the form :
\begin{equation}
H^{2}=\frac{1}{3M_{P}^{2}}\frac{V(\phi)}{\sqrt{1-\dot{\phi}^{2}}},
\end{equation}
\begin{equation}
\frac{\ddot{\phi}}{\sqrt{1-\dot{\phi}^{2}}}+3H\dot{\phi}=-\frac{(V_{,\phi}+\Gamma \dot{\phi} \sqrt{1-\dot{\phi}^{2}})}{V},
\end{equation}
where $V_{,{\phi}}$ represents partial derivative of $V$ with respect to $\phi$ and
\begin{equation}
{\dot{\rho}}_{\gamma}+4H\rho_{\gamma}=  \Gamma {\dot{\phi}}^2.
\end{equation}

In the inflationary epoch, it is assumed that $\rho_{\phi}~ \sim ~ V$ and $\rho_{\phi}~ > ~\rho_\gamma$.
Since in the slow-roll regime ${\dot{\phi}}^2\ll 1$ and $\ddot{\phi}\ll (3H+\frac{\Gamma}{V})\dot{\phi}$ and defining the dissipation rate $r$ as
$r\equiv\frac{\Gamma}{3HV}$,
the Friedmann equation and the equation of motion for the tachyon field in slow-roll regime respectively take the forms:
\begin{eqnarray}
H^2~&=&~\frac{V}{3 M_P^2}\\
3H(1+r){\dot{\phi}}~&=&~-\frac{V,_{\phi}}{V}.
\end{eqnarray}

For a quasi-stable radiation process during inflationary epoch, we need to satisfy the conditions $\dot{\rho_\gamma} \ll 4H\rho_\gamma$ and $\dot{\rho_\gamma}\ll \Gamma{\dot{\phi}}^2$. Hence in the slow roll regime, we can write
\begin{eqnarray}
\rho_\gamma~&=&~\frac{\Gamma {\dot{\phi}}^2}{4H}\\
&=& \frac{r~M_P^2}{4 (1+r)^2}\left(\frac{V,_{\phi}}{V}\right)^2\nonumber\\
&\equiv& \sigma {T_\gamma}^4\nonumber\\
&=& d (say)
\end{eqnarray}
where $T_\gamma$ is the radiation temperature and $\sigma$ is the Stephan-Boltzmann constant.

\section{Warm Inflation with Geometric Tachyon}

The radion field in case of one dimensional motion of a Dp-brane inside  a ring of NS $5$-branes is tachyonic \cite{PA1} which is the geometric tachyon.
The potential under this configuration of the geometric tachyon field is given by
\begin{eqnarray}
V(\phi)=V_{0}\cos\left(\frac{\phi}{\sqrt{k~l_{s}^{2}}}\right),~  && V_{0}=\frac{\tau_{3}R}{\sqrt{k~l_{s}^{2}}}
\end{eqnarray}
with the condition that
\begin{equation}
\sqrt{k ~l_s^{2}}\geq R
\end{equation}where $k$ is the number of NS5 branes in the ring of radious $R$ , $l_s$ is the string length scale and ${\tau}_3$ is the tension on the brane..

We restrict ourselves in the region of high-dissipation where $\Gamma \gg 3~H~V$ i.e $r\gg 1$.
Using the slow-roll approximation, the number of e-foldings ,~$N=\ln a$ from beginning to end of inflation then becomes
\begin{equation}
N={{\int}_{t_i}}^{t_f} Hdt=-{{\int}_{\phi_i}}^{\phi_f}\frac{3H^{2}r V(\phi)}{V'(\phi)}d\phi
=p\left[\cos\left(\frac{\phi}{\sqrt{k~l_{s}^{2}}}\right)-\cos\left(\frac{\phi_{f}}{\sqrt{k~l_{s}^{2}}}\right)\right]
=p[y-y_{f}].
\end{equation}
where $y=\cos(\frac{\phi}{\sqrt{kl_{s}^{2}}})$  and $p$ is a dimensionless parameter introduced via  $p=\frac{V_{0}}{4d}.$

The standard slow roll parameters $\varepsilon$ and $\eta$ in terms of the model parameters turn out to be
\begin{equation}
\varepsilon=-\frac{\dot{H}}{H^2}=-\frac{2 d}{V_{0}\cos(\frac{\phi}{\sqrt{kl_{s}^{2}}})}=\frac{1}{2py}
\end{equation}
and
\begin{equation}
\eta=-\frac{\ddot{H}}{H\dot{H}}=-\frac{3 M_{P}^{2}}{4 d k~l_{s}^{2}}\left[\frac{1}{2}\tan^{2}\frac{\phi}{\sqrt{k~l_{s}^{2}}}+1\right]\dot{\phi}^{2}=-\frac{y}{p(1-y^{2})}\left(1+\frac{1}{2}\frac{(1-y^{2})}{y^{2}}\right)
\end{equation}

We plot in Fig.1 the variation of $\eta$ and $\varepsilon$ with dimensionless field parameter $y$. As the plot shows, both the slow-roll parameters vary in a similar fashion except for large values of $y$. However, we point out that $\eta$ approaches unity faster than $\varepsilon$, the difference being visible  in the fourth decimal place of $y$ value only. Thus we consider the end of inflation to be marked by $\eta\simeq 1$

In terms of the dimensionless variable $p$ , this condition then reads as :
\begin{equation}
-\frac{1}{p}~=~\left[\frac{1}{2y}+\frac{y}{1-y^2}\right].
\end{equation}

\begin{center}
\begin{figure}[htbp]
\includegraphics[width=8cm]{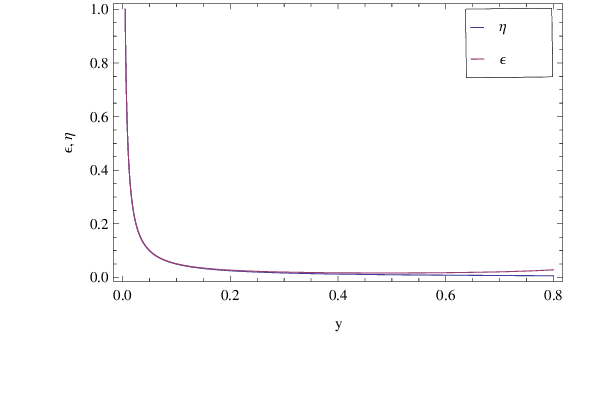}\\
\caption{Variation of the standard slow-roll parameters $\epsilon$ and $\eta$ with $y$}\label{ep-eta}
\end{figure}
\end{center}

For a given value of the parameter e.g $p=10^2$ , we can have the required number of e-foldings  $N\sim 60$ before the end of inflation provided the tachyon field  starts rolling from an initial value of  $y~=~0.60500025$.

With this background results, we now discuss the perturbative spectrum in the next subsection.

\subsection{Perturbation Spectrum}

One of the basic features of warm inflationary scenario is that the production of radiation during inflation  influences the seeds of density fluctuations and in high dissipative regime, the density perturbation is expressible as \cite{PA2}
\begin{equation}
\delta_{H}=\frac{16 \pi}{5}\frac{M_{P}^{2}\exp(-\bar{\zeta}(\phi))}{(\ln(V)),_{\phi}} \delta{\phi},
\end{equation}
where,
\begin{eqnarray}
\bar{\zeta}(\phi)&=&-\int\left[\frac{1}{3Hr}(\frac{\Gamma}{V}),_{\phi}
+\frac{9}{8}(1-\frac{(\ln(\Gamma)),_{\phi}(\ln(V)),_{\phi}}{36 H^{2}r})\right] d\phi \\ \nonumber
&=&-\int\left[2\frac{V''}{V'}-\frac{3V'}{8V}-\frac{3dV''}{4VV'}+\frac{3dV'}{16V^{2}}\right] d\phi
\end{eqnarray}

From the above expression of density perturbation one can derive the relevant expression for various cosmological observable \cite{Herrera}. We present below the observable in context of the  present model.

\begin{flushleft}
\bf{(a) Spectral Index : }
\end{flushleft}
This quantity is found to be
\begin{eqnarray}
n_{s}&\equiv&1+\frac{d~ln\delta_H^2}{d~ln k}\\ \nonumber
 &=&1-\left[\frac{3\eta}{2}
+\epsilon\left(\frac{2V}{V,_{\phi}}(2\bar{\zeta}(\phi)-\frac{r_{\phi}}{4r})-\frac{5}{2}\right)\right]\\ \nonumber
&=&~\frac{3d^{2}}{2V^{2}}+\frac{3d}{V}+1-\frac{6d^{2}V''}{V(V')^{2}}
+\frac{12V''d}{(V')^{2}}\\ \nonumber
\end{eqnarray}
In terms of dimensionless variables $p$ and $y$, this reads as
\begin{equation}
n_{s}=1+\frac{3}{32p^{2}y^{2}}+\frac{3}{4py}+\frac{3}{8p^{2}(1-y^{2})}-\frac{3y}{p(1-y^{2})}.
 \end{equation}

  \begin{figure}[htbp]
  \centering
  \includegraphics[width=8cm]{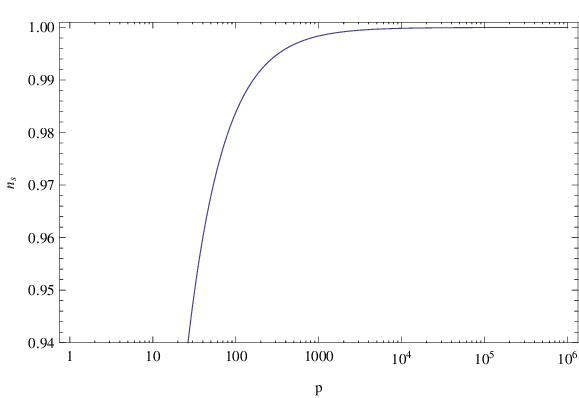}\\
  \caption{Variation of Spectral Index with $p$}\label{spectral}
\end{figure}

 In the above figure the variation of spectral index with the model parameter $p$ has been presented. We observe that at $p=100$, the spectral index is  $n_s=0.983853$. Thus the model produces the expected scale invariance of the power spectrum.

\begin{flushleft}
\bf{(b) Running spectral index :}
\end{flushleft}
This observable is defined and turns out in the present model as follows :
 \begin{eqnarray}
 \alpha_{s}&\equiv&\frac{d~n_s}{d~ln~k}\\ \nonumber &=&-\frac{2V\epsilon}{V,_{\phi}}\left[\frac{2\eta,_{\phi}}{2}
 +\frac{\epsilon_{\phi}}{\varepsilon}\left(n_{s}-1+\frac{3\eta}{2}\right)
 +2\epsilon\left((\frac{V}{V,_{\phi}}),_{\phi}(2\bar{\zeta},_{\phi}
 -\frac{(\ln(r)),_{\phi}}{4})
 +(\frac{V}{V,_{\phi}})(2\bar{\zeta},_{\phi\phi}-\frac{(\ln(r)),_{\phi\phi}}{4})\right)\right]\\ \nonumber
 &=&-\frac{12d^{2}}{V^{3}(V')^{4}}[(d+V)(V')^{4}-2dV(V')^{2}V''+4V^{2}[-d+2V](V'')^{2}
 +2[d-2V]V^{2}V'V''']\\ \nonumber
 \end{eqnarray}
 In terms of the model parameters, $\alpha_{s}$ reads as
 \begin{equation}
 \alpha_{s}=-\left(\frac{3}{16 p^{3}y^{3}}+\frac{3}{4p^{2}y^{2}}+\frac{3}{2p^{2}(1-y^{2})^{2}}-\frac{3y}{4p^{3}(1-y^{2})^{2}}+\frac{3(-1+2y^{2})}{2p^{2}(1-y^{2})^{2}}\right).
 \end{equation}
We plot below the variation of $\alpha_s$ with $p$ and find that for $ p=30$, $\alpha_{s}$ becomes -0.00530187. Thus this observable also remains consistent with the observation.

\begin{figure}[htbp]
  \centering
  \includegraphics[width=8cm]{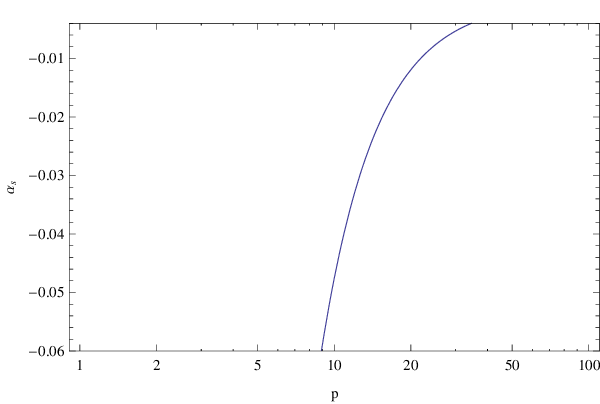}\\
  \caption{Variation of Running spectral index with model parameter $p$}\label{alpha}
\end{figure}
\pagebreak

\begin{flushleft}
\bf{(c) Power Spectrum : }
\end{flushleft}
The Power-spectrum which is proportional to the square of density perturbation, in the present case turns out to be
 \begin{eqnarray}
 P&\equiv&\frac{25}{4}\delta_{H}^{2}\\ \nonumber
  &=&\frac{1}{4\pi^{2}}\left[\frac{1}{\sigma d}\frac{V^{6}}{(V')^{4}}\right]^{\frac{1}{4}}\exp(-2\bar{\zeta}(\phi))\\ \nonumber
  &=&\frac{1}{4\pi^{2}}\left(\frac{fp^{2}y^{6}}{(1-y^{2})^{2}}\right)^{\frac{1}{4}}\times \\ \nonumber
  & &\exp\left[-\frac{3}{4}\log(y)
+\frac{3}{8p}\log(\cos(\frac{1}{2}\arccos(y)))-\log(1-(\cos(\frac{1}{2}\arccos(y)))^{2})^{\frac{1}{2}}\frac{3}{8p}
+4\log(1-y^{2})^{\frac{1}{2}}-\frac{3}{32py}\right]
 \end{eqnarray}
 where $f=\frac{16k^{2}l_{s}^{4}}{\sigma}$ is a dimensionless constant . From the observational constraint on the power-spectrum \cite{COBE} i.e $ P=2\times10^{-9}$, we estimate the value of this constant to be
$f=2.69145\times10^{-31}$ for $p=100$ and $y=0.605$.

\begin{flushleft}
\bf{(d) Tensor to scalar ratio : }
\end{flushleft}
One more cosmological observable which is tensor to scalar ration is supposed to play crucial role in gravity wave production and is defined as :
\begin{equation}
R\equiv\left[\frac{A_{g}^{2}}{P}\right]_{k=k_{0}}
\end{equation}
where $A_{g}^{2}$ is the tensor spectrum . The tensor spectrum is given by,
\begin{equation}
 A_{g}^{2}=\frac{H^{2}}{2\pi^{2}M_{P}^{2}}\left[\coth\left(\frac{k}{2T}\right)\right]_{k=k_{0}}
\end{equation}
This gives :
\begin{eqnarray}
R&=&\frac{2}{3M_{P}^{4}}\left[\frac{\sigma d (V')^{4}}{V^{2}}\right]^{\frac{1}{4}}
\exp(2\bar{\zeta}(\phi))\left[\coth\left(\frac{k}{2T}\right)\right]_{k=k_{0}} \\ \nonumber
&=&\frac{8d}{3}\left(\frac{p^{2}(1-y^{2)}}{f y^{2}}\right)^{\frac{1}{4}}\times \\ \nonumber
& &\exp\left[+\frac{3}{4}\log(y)
-\frac{3}{8p}\log(\cos(\frac{1}{2}\arccos(y)))+\log(1-(\cos(\frac{1}{2}\arccos(y)))^{2})^{\frac{1}{2}}\frac{3}{8p}
- 4\log(1-y^{2})+\frac{3}{32py}\right]
\end{eqnarray}
Using $T_\gamma \cong 0.24\times 10^{16} GeV $ and $k_0=0.002Mpc^{-1}$ \cite{PA2}, the value of $R$ is obtained as $R= 0.0091596$ which is well within the observational limit. FIG.4 shows the variation of $R$ with respect to the model parameter $p$.

\begin{figure}[httb]
  \centering
  \includegraphics[width=8cm]{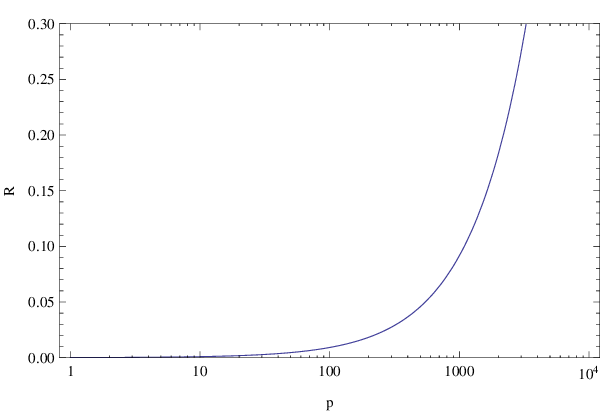}\\
  \caption{Variation of $R$ with the parameter $p$}\label{ar}
\end{figure}

\subsection{Non-Gaussianity}

Single field inflation models broadly predict Gaussian primordial density fluctuations. When second order effects are taken into account, there are corrections to this general prediction.There is a resulting non-Gaussian signal in CMB, magnitude of which depends on the self-interaction of the inflaton field \cite{GB}.

It is possible to write an expression for the non-gaussianity parameter $f_{NL}$ in strong dissipative regime \cite{Kamali} as

\begin{equation}
f_{NL}=-\frac{5}{3}\left(\frac{\dot{\phi}}{H}\right)\left(\frac{1}{H}\log(\frac{k_{f}}{H})\right)\left[\frac{V'''(\phi)}{\Gamma}+\frac{(2k_{f})^{2}V'(\phi)}{\Gamma}\right]
\end{equation}

For the model parameters $y$, $p$ and $d$ , $f_{NL}$ becomes
\begin{equation}
f_{NL}=\frac{10[\frac{2}{y^{2}}+\frac{p}{y}-\frac{4}{1-y^{2}}]\log[\frac{\sqrt{3}\sqrt{1-y^{2}}}{2\sqrt{d}}]}{3p^{2}}
\end{equation}
When $p=100$, $y=.605$,$d=10^{-12}M_{P}^{4}$, $f_{NL}$ becomes 0.736917 which is well within the bound observed by WMAP7 \cite{WMAP}.

\subsection{Constraints on the number of branes}

From our analysis,we have the slow roll parameter as
\begin{equation}
\epsilon=\frac{1}{2 p y}
\end{equation}
where $p=\frac{V_{0}}{4d}$ and $y=\cos[\frac{\phi}{\sqrt{kl_{s}^{2}}}].$
Taking the condition $\epsilon<<1:$, one finds that the models parameters are constrained by $\frac{4d}{2V_{0}y}<<1$.

From Eq.9   we have $V_{0}=\frac{\tau_{3}R}{\sqrt{kl_{s}^{2}}}$ and taking note of the fact that brane tension $\tau_{3}$ is related with the string coupling $g_{s}$ via the relation $\tau_{3}=\frac{M_{s}^{4}}{(2\pi)^{3}g_{s}}$, ($M_{s}$ being the string mass scale) and $d=10^{-12}M_{P}^{4}$, we get,
\begin{equation}
\frac{2\times 10^{-12}M_{P}^{4}\sqrt{kl_{s}^{2}}(2\pi)^{3}g_{s}}{y M_{s}^{4}R}<<1.
\end{equation}
But the tachyon potential has been obtained with the approximation \cite{PA1}
\begin{equation}
\sqrt{k}~l_{s}>>R ,
\end{equation}
Taking these into consideration, for the model parameter for $p=10^2$ number of background branes have the constraint
\begin{equation}
k<<10^{6}.
\end{equation}

Again, the parameter $p$ in our analysis is defined as $p=\frac{V_{0}}{4d}$. Using typical values of the parameters used in this analysis,we get a relation $\frac{R}{\sqrt{k}l_{s}g_{s}}=4\times 10^{-4}$.
Taking the value of R to be $10^{-3}l_{s}$ and using the fact that for the effective theory of tachyon to be valid one needs to work in a weak coupling regime [$g_{s}<<1$],finally we get the constraint on k as
\begin{equation}
k>>6.25.
\end{equation}

 So, for the present model to work,the number of branes $10^6 \gg k\gg 6 $. It may be noted that the minimum number of branes required here is less than that needed for the cold case [$k>>10^{4}$], which is an interesting result.

\section{Discussion}

In this work we investigated the viability of warm tachyonic inflationary scenario driven by geometric tachyon. Warm Inflationary scenarios are free from the so called $\eta$-problem with dissipative effects slowing down the inflaton's motion and producing radiation during inflation avoiding the need of reheating the system.Here the particle production is sourced by the rolling tachyon field itself so that the radiation is not completely diluted away in the slow-roll regime.
Introduction of the dissipative effect in the equation of motion of the tachyon field modifies the definition of slow roll parameters as well as all other cosmological observables. We have computed the spectral index $n_s$, Running of spectral index $\alpha_s$, Power spectrum $P$ and tensor to scalar ratio $R$ in terms of our model parameters.Interestingly in this model, the string lengh scale along with the number of NS $5$-branes in the background plays a crucial role in behaviour of the cosmological observables .And we observe that to get all the cosmological observables within the experimental bounds \cite{WMAP}, the  number of NS $5$ branes required to produce the necessary background, in which the $D3$ brane moves, gets drastically reduced compared to the cold scenario.
Hence a successful warm inflation can be realized even in the background produced by ten NS $5$ branes. In cold inflationary scenario, a successful model of inflation driven by geometric tachyon need presence of  approximately $10^4$ NS $5$ branes to produce the necessary background.
In both weak and strong regimes of Warm-inflation, fluctuations in the radiation are transferred to the inflaton and become the primary source of density fluctuations.An important feature of warm inflation is that a significant amount of non-gaussianities get produced whilst the density fluctuations are still sub-horizon scale.Non-linear evolution during the inflationary era can result in non-gaussianities appearing in the primordial fluctuations.We estimated the amount of non-gaussianity in our allowed range of model parameters and found that it is quite insignificant
to be detected in the scope of present day observations.
\begin{center}
\bf{Acknowledgements}
\end{center}
 Financial support from DAE Project grant 2009/24/35 (BRNS) is acknowledged. We also acknowledge Sudhakar Panda for valuable discussions.

%%%%%%%%%%%%%%%%%%

\end{document}